
\input phyzzx

\def\square{\hbox{{$\sqcup$}\llap{$\sqcap$}}} 
\def\part{\partial}

\def\dD{\hbox{D}}
\def\ee{\hbox{e}}
\def\eg{{\it e.g.}, }
\REF\vj{ J.D. Cohn and V. Periwal, IAS/Fermilab preprint IASSNS-HEP-92-18,
Fermilab-92/126-T, Nucl. Phys. B to appear}
\REF\ps{ J. Polchinski and A. Strominger, {\sl Phys. Rev. Lett.} {\bf %
67} (1991) 1681}
\REF\fy{ P.H. Frampton and M. Yamaguchi, Univ. of North Carolina %
preprint IFP-420-UNC (1992)}
\REF\wz{ D.J. Wallace and R.K.P. Zia, {\sl Phys. Rev. Lett.} {\bf 43} %
(1979) 808}
\REF\dkw{ H.W. Diehl, D.M. Kroll and H. Wagner, {\sl Z. Physik }%
{\bf B36} (1980) 329}
\REF\ml{ M. L\"uscher, {\sl Nucl. Phys.} {\bf B180} (1981) 317}
\REF\ll{ S.C. Lin and M.J. Lowe, {\sl J. Phys.} {\bf A16} (1983) 347}
\REF\no{ H.B. Nielsen and P. Olesen, {\sl Nucl. Phys.} {\bf B61} (1973) 45}
\REF\da{R. L. Davis and E.P.S. Shellard, {\sl Phys. Lett. }{\bf 214B} (1988)
219 }
\REF\tur{N. Turok, ``Phase Transitions as the Origin of Large Scale
Structure in the Universe'' in TASI 1988 proceedings, {\it Particles, Strings
and Supernovae}, Eds. A. Jevicki and C.-I. Tan (World Scientific:
1989, Singapore)}
\REF\rg{ R. Gregory, {\sl Phys. Rev. }{\bf D43} (1991) 520}
\REF\gs{ J.-L. Gervais and B. Sakita, {\sl Phys. Rev.} {\bf D11} %
(1975) 2943; {\bf D12} (1975) 1038}
\REF\cl{ N.H. Christ and T.D. Lee, {\sl Phys. Rev.} {\bf D12} (1975) %
1606}
\REF\gp{ E. Gildener and A. Patrascioiu, {\sl Phys. Rev.} {\bf D16} %
(1977) 423}
\REF\raj{R. Rajaraman, {\it Solitons and Instantons} %
(North Holland: 1987, The Netherlands)}
\REF\hp{ J. Hughes and J. Polchinski, %
{\sl Nucl. Phys.} {\bf B278} (1986) 147}
\REF\ol{P. Olesen, {\sl Phys. Lett.} {\bf 160B} (1985) 144}
\REF\poly{ A. Polyakov, {\sl Nucl. Phys.} {\bf B120} (1977) 429}
\REF\regge{ See for example, M. Perl, {\it High Energy Hadron Physics } %
(Wiley Interscience: New York) 1974, p. 390}
\REF\wils{K. G. Wilson, {\sl Phys. Rev.} {\bf D10} (1974) 2445}
\REF\wein{D. Weingarten, {\sl Phys. Lett.} {\bf 90B} (1980) 285}
\REF\largen{G. 't Hooft, {\sl Nucl. Phys.} {\bf B72} (1974) 461}
\REF\polqcd{ J. Polchinski, {\sl Phys. Rev. Lett.} {\bf 68} (1992) 1267}
\REF\polstring{ J. Polchinski, Univ. of Texas preprint UTTG-16-92 (1992)}
\REF\dl{ D. C. Lewellen, ITP preprint NSF-ITP-91-105 (1991)}

\def\pc{\phi_{\rm cl}}
\def\dd{\hbox{d}}
\def\de{\delta}
\def\ha{\hat a}
\def\zft{z - f(t,y)}

\def\part{\partial}

\def\sql{\sqrt{\lambda}}

\footline={\hss\tenrm\folio\hss}
\nopubblock
\titlepage
\rightline{Fermilab 92/293-C}
\vskip-8truept
\rightline{iassns-hep-92-73}
\vskip-8truept
\rightline{hep-th/9210156}
\title{Quantizing Effective Strings\foot{Lecture given by
J.D.C. at TASI summer school in Boulder, CO, 1992, to appear in the
proceedings}}
\author{{\rm J.D. Cohn}\foot{jdcohn@fnal.fnal.gov}}
\address{Fermilab, MS 106, P.O. Box 500 \break
Batavia, Illinois 60510}
\andauthor{{\rm Vipul Periwal}\foot{vipul@guinness.ias.edu}}
\address{The Institute for Advanced Study \break
Princeton, New Jersey 08540-4920}
\abstract
\baselineskip=14truept
This talk reviewed the theory of effective strings, with
particular emphasis on the manner in which Lorentz invariance
is represented.  The quantum properties of an example of an effective
string are derived from the underlying field theory.
A comparison is made with what one would expect if one assumed that
quantum effective strings were governed by fundamental string actions
such as the Nambu-Goto or the Polyakov actions.  It is shown that
the requirements on dimensions for consistent quantizations
of fundamental strings imply no contradictions for effective strings.

\endpage
\pagenumber=1
\baselineskip=14truept
\parskip=1truept
There are several contexts when the physics of
many particle systems, at some length scale or for some range of
parameters, is simplest understood in
terms of effective stringlike excitations.
In some of these cases, one has a map from an underlying field
theory to stringlike variables, and
one can derive properties of the resulting `string theory'
from the field theory.
This kind of `explicit' effective string is the main subject of this
review, which will follow for the most part the presentation of [\vj].
As is well known, fundamental strings have critical dimensions.  One
way of understanding these special dimensions is to realize that,
unless additional degrees of freedom are incorporated,
quantum fundamental strings, described by either the Polyakov or the
Nambu-Goto actions, are only Lorentz invariant in these
dimensions.  The motivation for [\vj], which calculated the induced
Lorentz transformations
in a theory of effective strings from the underlying field theory, was to
figure out how effective strings evade the problem of Lorentz
non-invariance.

The resolution we shall find of this apparent paradox is very simple: Lorentz
transformations of the effective string have a geometric universal term of
dimension $-1$ that leads to a mixing between the dimension 2 term
in the action that describes long distance physics on the string
worldsheet, and an irrelevant dimension 4 term that does not affect
long distance physics.  The key to this structure is the
fact that effective strings have a length scale---they are thick.
Thus one has stringlike infrared logarithmic divergences in correlations,
just as in fundamental strings, but the fact that there is a length
scale alters ultraviolet properties.
The short distance operator product algebra used to compute the
conformal anomaly (which is at the heart of the Lorentz non-invariance
of fundamental strings) is irrelevant for the Lorentz
invariance of effective strings.

The problem of the quantization of fundamental strings in dimensions
other than their critical dimensions, or of effective strings with conformal
invariance (if such strings exist), is not addressed here.
For a suggestion, see [\ps], and for further study of this
suggestion, see [\fy].

The rest of the talk is as follows.
After listing some theories with explicit stringlike variables,
an example in 2+1 dimensions is discussed
in detail.  This system has been well studied in the literature, {\it e.g.},
[\wz,\dkw,\ml,\ll].  It is shown how the quantization of
the underlying field theory induces one in the effective string theory.
In particular, the induced Lorentz transformations are explicitly
derived.  Comparisons with fundamental
strings are made, including a reminder of the comparison made
in the classic paper by Nielsen and Olesen[\no].   Some
`implicit' effective strings are mentioned for contrast at the end.

An example of an explicit string
is a domain `string' in 2+1 dimensions, in an Ising model,
separating regions on a plane
where spins point up from regions where the spins point down.
In each region, the order parameter, the magnetization, has a definite
sign, while on the string the magnetization goes to zero.  Another example
is a flux tube in 3+1 dimensions,
{\it e.g.} a Nielsen-Olesen string in
the Abelian Higgs model[\no] .  Physical flux tubes include
type II superconductors and cosmic strings.  The position of the string
is specified by where the order parameter goes to zero (the magnetization
for the Ising case, the expectation value of the Higgs field for the
gauge case).
Vortex rings, for example in hydrodynamics,
can also be described as a string theory, sweeping out a two dimensional
world sheet.  The antisymmetric tensor coupling in the string world sheet
action is related to the vorticity[\da].
Some of these strings have externally preferred directions (a flux tube in a
superconductor has an external magnetic field) or are self avoiding.

For explicit effective strings, as will be shown in detail,
it is possible to rewrite a functional integral $\int\dD\phi\ \ee^{iS}$
about a stringlike background in terms of modes ($f$) that correspond to
fluctuations of the string (referred to as string configurations in
the following) and other modes ($a$)  that are separated from the string
configurations by a mass gap.  These modes excite the internal structure
of the effective string.  They may be integrated out for the purposes of
studying the long distance properties of the effective string.
Thus one has
$$\int\dD\phi\ \ee^{iS(\phi)}\bigg|_{\rm about\ a\ string\ solution}
= \int\dD f\dD a\ \ee^{iS(f,a)}
= \int\dD f\ \ee^{iS_{\rm eff}(f)} \ . $$
So an effective string appears as a quantized string theory,
a sum over different string configurations
weighted by some effective string action $S_{\rm eff}.$

We will be interested in comparing $S_{\rm eff}$
with natural geometric actions that one would consider for
structureless `fundamental'
string theories, \eg the Nambu-Goto action, which is just the
area in spacetime of the string worldsheet,
$
S \sim \int \sqrt {- \det \partial_\mu X^i \partial_\nu X_i} \; .
$
An early comparison for a string in the Abelian
Higgs model is [\no], more
recent comparisons include the 1988 TASI lectures[\tur] which focuses
on cosmic strings.
In the long wavelength limit for
the effective string, and for the structureless string in light cone gauge,
the action is proportional to
$(\partial_\mu f^i)^2$ where the $f^i$ are the transverse coordinates of the
string in spacetime.
(That is, $i = 1, \dots D-2$, $f^{D-1}$ lies along the string
and $f^{D}$ is time.)  In addition
an explicit effective string has a scale, a width,
$m^{-1}$.  Its field theory is nonrenormalizable.
There are corrections to the purely geometric action
which depend upon the short
distance physics of the underlying field theory:
$$
S \sim \int (\partial_\mu f^i)^2 + b [(\partial_\mu f^i)^2]^2 + \dots
$$
Here $\partial \sim m^{-1}$, {\it i.e.}
the expansion is a long wavelength expansion.

One way to quantize the Nambu-Goto string
is to use the Polyakov action, whose classical equations
of motion agree with those of Nambu-Goto.  The Polyakov action has a
larger invariance, the freedom of Weyl rescalings of the intrinsic
metric.  The conformal anomaly implies that this classical gauge
invariance is not a symmetry of the quantum theory.
In conformal gauge, this anomaly means
one needs either $D=26$ or the Weyl degree of freedom
does not decouple.

Another approach to the Nambu-Goto string is to
attempt quantization in light cone gauge.  There one finds that
unless $D=26,$ Lorentz invariance is lost (the anomaly in the
Lorentz algebra vanishes for $D=2,3$ but there are still
problems with interactions).
There is no known consistent quantization of Nambu-Goto strings
in any dimension between $3$ and $25,$ which is one of the reasons
for interest in how effective strings evade the Nambu-Goto
string's problems.

One procedure for studying effective strings is to take
$S_{\rm eff}$ and write the terms relevant for long distance physics in
geometric form.  Quantizing the resulting geometric classical
action results in the usual Nambu-Goto string.
However, $\int\dD f\ \ee^{iS_{\rm eff}(f)},$ induced from the underlying
field theory, already defines
a quantum string theory, and it is the properties of this theory
which will be discussed in the following.  We shall see that
there is some universal stringlike behavior even
though conformal invariance, a usual characteristic of
fundamental strings, does not appear.
The steps involved in going from a theory with a stringlike
solution to the equations of motion to an effective
string action are:  (1) to look
at fluctuations around the string background; (2)
to introduce a string coordinate,
integrate out the massive excitations (possible since
there is a mass gap between internal
excitations and the zero mass excitations which arise due to broken
translational symmetry);
and then (3) to
use the field theory quantization
to find the string quantization (\eg the Lorentz
transformations).  Another possible route to writing an effective string
theory is to assume a string solution and
expand in the width of the string[\rg].

The specific example in the following is a domain string in $2+1$ dimensions.
The Lagrangian is
$$ {\cal L} = {1\over 2} \left[ \part_r\phi\part^r\phi
- \lambda (\phi^2-{m^2\over \lambda})^2\right],$$
the metric $\eta_{ij}$
has signature $+--,$ and the coordinates are $x^r\equiv(t,y,z)
\equiv(y^\mu,z).$
This is an Ising-ferromagnetlike system.  Some of the classic references
for the string description of this theory (called the `drumhead model'
in some contexts)
are [\wz,\dkw,\ml,\ll] and references therein.  This
example is used here because it has been so thoroughly studied that
many of the calculations can be done explicitly.

There are two minima of the potential, $ \phi = { \pm m /\sqrt{\lambda}}$.
The equation of motion is
$$\part^2\phi +2\lambda(\phi^2 -{m^2\over \lambda})\phi = 0$$
which admits a solution interpolating between the two minima.
Choosing this interpolation to take place along the $z$ direction,
the solution can be written as
\def\sech{\hbox{sech}}
$$\pc(z) \equiv\ {m \over \sqrt \lambda}  \tanh mz.$$  This
describes a domain string at $z=0$.
Other solutions with the same boundary conditions exist,
corresponding to multiple string configurations,
but  will not be considered.  It should
be kept in mind that a long straight string is unstable:
as will be seen, its motion is described by two
dimensional massless bosonic fields,
which have an infrared logarithmic divergence in their correlators.
To define things carefully appropriate boundary conditions, or a small
mass term, should be included in the discussion.

The field theory in this background is
$$
{\cal L}(\pc + \tilde{\xi}) =
{\cal L} (\pc) - \tilde {\xi} (\square + \Omega(z)) \tilde{\xi}
+ O( \tilde{\xi}^3)$$
where
$\square = \part_t^2 - \part_y^2$ and $\Omega =
-(\part_z - 2 \sqrt{\lambda} \pc(z))(\part_z + 2 \sqrt{\lambda} \pc(z))$
The explicit form of $\pc$ was used to rewrite ${\cal L}$.
The spectrum and eigenfunctions of the quadratic fluctuation operators
around this background are known.  For $\square$ the eigenfunctions
are plane waves in $(t,y)$:  $e^{i( \omega t + k_y y)}$.
\hfuzz=6truept
For $\Omega$:
$$\eqalign{{\rm eigenvalue\; \; \; } & {\rm eigenfunction} \cr
0 \quad & \psi_0 =
{\sqrt{3m}\over 2} { }\sech^2(mz)
\equiv{\sqrt{3\lambda\over 4m^3} } \pc '(z) ,\cr
 ***  \quad & {\rm gap} \cr
3m^2  \quad & \psi_1 = \sqrt{3m\over 2}  \sech(mz)\tanh(mz),\cr
k^2+4m^2 \quad & \psi_k = {{\sqrt m\exp(ikz)}\over
{\sqrt{k^4+5k^2m^2+4m^2}}}
\left[3\tanh^2(mz)-{3ik\over m}\tanh(mz)-{k^2 \over m^2}-1\right]\cr
& -\infty < k < \infty  \cr}$$
These modes have the following physical significance:
\item{(a)} The zero mode $ \psi_0 \propto \part_z \pc$
is the Nambu-Goldstone boson corresponding
to the translation invariance broken by the string,
$\pc(z + \beta) = \pc(z) + 4 m^3/3 \psi_0(z) + \cdots$
\item{(b)} The mode with mass $\sqrt{3} m$ is also localized
on the string and is
referred to as the kink `excitation' in the literature.
It corresponds to a squeezing of the string:
$\pc(z(1+\beta)) = \pc(z) + \beta z \part_z \pc (z)$.  For
this specific case,
the normalized overlap of $z\dd\pc/\dd z$ and $\psi_1$
is $\pi\sqrt3/\sqrt{8\pi^2-48} \approx 0\cdot978.$
\item{(c)} A continuum starting at mass $2m,$ with $k$ taking
arbitrary real
values.  These extended
modes are the counterparts of the spectrum obtained when
expanding about  a homogeneous background $\pc(x) = \pm m/\sql$.

Naively in quantizing this system one would assume that fluctuations
$\tilde{\xi}$ around the background are small.  This is not true of
the zero mode, which describes a fluctuation with no damping.
One could have put the domain string anywhere.  To quantize this system
correctly one treats the zero mode exactly by introducing
an (implicit) collective coordinate.  As a result of introducing the collective
coordinate the position of the string $f(t,y)$ will be introduced into the
classical solution,
$\pc(z) \rightarrow \pc(z-f(t,y))$.  All $f(t,y)$ will
be integrated over, and the
zero mode in $\tilde{\xi}$ will be projected out.

The introduction of collective coordinates is done
by analogy with Fadeev-Popov ghosts.  Collective coordinates
for solitons are due to [\gs], the
method applied to implicit collective coordinates is due to [\cl,\gp].
A pedagogical treatment of the general idea can be found in [\raj].
Use the identity
$$
\int  \dD f(t,y)\ \delta(g(f)) |{\de g \over \de f}| = 1
$$
inside the functional integral for $\phi$:
$
\int \dD \phi\  \ee^{iS(\phi)} \; .
$
For the case here choose
$
g(f) = \int \dd z \part_z \pc(\zft) \phi(z) \; .
$
In the functional integral, $\de (g(f))$ projects out the zero mode,
$\tilde{\xi} \rightarrow \xi$ and introduces an integral over
the position of the
string $f(t,y)$ into the measure.

The explicit form of $\pc$ describes one kink, so it is
being assumed that there are no overhangs, that the string is only at one
$z$ position.  This means multikink solutions are neglected.
These are down by $e^{-mR}$ in the functional integral, where $R$ is the
length of the string, when the kinks are widely separated.
The Jacobian,
$$
| {\de g \over \de f} | = \int \dd z \part_z \pc(\zft) \part_z \phi(z)
\equiv \Delta (\xi)$$
is independent of $f(t,y)$.  It can be set to one
using dimensional regularization[\wz,\ll].

To introduce the collective coordinate $f(t,y)$
into the rest of the action write
$$\phi(t,y,z) \equiv \pc\left(z-f(t,y)\right) +
\xi\left(t,y,z-f(t,y)\right),$$ and plug in to the functional
integral to get
$$
Z = \int \dD \xi(t,y,z) \dD f(t,y)\ \ee^{iS(\pc + \xi)}\  \Delta (\xi )
+ {\rm multistring \; \; configurations} \; .
$$
All the $f$ dependence is in the action $S$ only, not in $\Delta$.
The measure for
$f$ is ultralocal, depending only on the value of $f$ at a given point,
not upon derivatives of $f$.  The measure for $\xi$ is that implied by
its decomposition in terms of eigenmodes of ${\Omega}$;
$$
\xi(t,y,z+\alpha) =
a_1 (t,y)\ \psi_1(z+ \alpha) + \int \dd k\ a_k(t,y)\ \psi_k(z+\alpha)
$$
and the coefficient of $\psi_0$ has been set to zero by the delta
function in the functional integral. The integral
over $k$ is schematic, it is not necessary to be precise since
loop effects will not be considered.

\def\ho{\hat\Omega}
Substituting, scaling out $m,\lambda$ and making all the $f$ dependence
explicit,
the action becomes
$$\eqalign{S = - {m\over \lambda}\int \dd^3x \bigg\{ {\pc '}^2 -
{1\over 2} \part_\mu f\part^\mu f\left[\pc ' + \xi ' \right]^2
&-{1\over 2} \xi\left(-\part_\mu\part^\mu - \ho \right)\xi
\cr
&+2\pc\xi^3 + {1\over 2} \xi^4 + \part_\mu\xi\part^\mu
f \xi ' \bigg\} .\cr}$$
The operator $\ho$ has no zero modes and primes denote $\part_z$.
This action describes the two dimensional field $f(t,y)$ interacting
with the massive three dimensional field $\xi (t,y,z)$, corresponding to
the rest of the degrees of freedom in this background.
This procedure also is
the one used for studying field theories in soliton backgrounds.

Here the goal is to consider the effective field theory of the domain string
with position at
$$
X^2  = f(t,y),\qquad X^1  = y ,\qquad X^0  = t.
$$
As a two dimensional field theory this is
a string (in a certain gauge) interacting with massive fields $a_k(t,y)$.
To get the theory of the string alone, eliminate the other degrees of
freedom, $\xi$, by using the equations of motion.
(This is leading order in $\hbar$ and corresponds to a saddle point
expansion for the heavy field $\xi$ in the
functional integral):
$$\eqalign{\xi = - \ho^{-1} \pc '' (\part f)^2 &+
 \ho^{-1} \left[\part_z^2 (\part f)^2
- 6\pc \ho^{-1} \pc '' (\part f)^2 \right]\ho^{-1} \pc '' (\part f)^2
\cr &+ \ho^{-2}\pc '' \part_\mu\part^\mu(\part f)^2 + \cdots\ . \cr}$$
This is a long wavelength expansion ($\part \sim m^{-1}$).
Since $\ho$ has no zero mode it is invertible.

The action for one domain string, with the massive
fields integrated out is then
$$\eqalign{ S(f) = -\int \dd t \dd y
&\bigg[ {1 \over 2 \pi \alpha '}(1 - {1 \over 2} (\part f)^2
- {1 \over 8} ((\part f)^2)^2 - {1 \over 16} ((\part f)^2)^3 + \cdots )
  \cr
& +  b (\part f)^2 \square (\part f)^2 + \cdots  \bigg] \; . \cr
{1 \over 2 \pi \alpha '} & = {m\over\lambda}\int \dd z (\pc ')^2, \qquad
b = {m\over 8\lambda} \int \dd z (z \pc ')^2 . \cr }
$$
As $f$ is a Nambu-Goldstone boson, coming from the
breaking of translation invariance, only derivatives of it appear.
The first three terms in the derivative expansion (including the constant)
come from the kinetic terms in the original action.
They are independent of the details of the potential except for the overall
factor of $\alpha '$.
The top line in $S$
(as was shown by [\dkw]) is the Nambu-Goto action for the string
$S_{N-G} = \sqrt {1 - (\part f)^2} = \sqrt{-\det \; h_{\mu\nu}}$.
The induced metric on the string world sheet $h_{\mu \nu}$ is
$h_{\mu\nu} = {\part X^i \over \part y^\mu}{\part X_i \over \part y^\nu}
= \eta_{\mu\nu} - \part_\mu f \part_\nu f$.
The second line is partially the intrinsic curvature, but also has a
contribution
that is not in any obvious way geometrical.   Since the coefficient of
this term is not universal, this is not surprising.
This expansion is up to $O(\part^8, \hbar,{\rm boundary \; terms})$.
To include higher order $\hbar$ effects, one needs to include loop effects
in the underlying field theory and
then find the new solution to the equations of motion and expand around
it.   If regularization schemes other than dimensional regularization
are used for the Jacobian $\Delta$, it may also contribute at order $\hbar$.

The quantization of the underlying field theory induces a quantization
of the string theory.  For instance, to find the Lorentz transformations
in the theory,
start with the transformations in the $\phi$ field theory.  This
example is worked out in detail in [\vj].
The canonical Lorentz generators are
$ M_{rs} \equiv  \int \dd y\dd z \left[j_{0r}x_s - j_{0s}x_r\right].$
where
$j_{rs} \equiv -\eta_{rs} {\cal L} + \part_r\phi\part_s\phi$
are the translation currents.
Upon quantization, $M_{rs}$ becomes an operator:  in the usual way
$ P_{\phi} = \part_0 \phi = - i \de/ \de \phi $ where
$$[P_{\phi}(y,z), \phi(y', z')]_{\rm e.t.} = - i \de (y-y') \de (z - z').$$
This induces a quantization of the string coordinate,
$[P_f(y), f(y')]_{\rm e.t.} = - i \de (y-y')$.
One can rewrite $P_{\phi}$ using the chain rule (using the components
of $\xi$, the $a_k$, and $\langle g|h \rangle = \int \dd z g(z) h(z)$):
$$
P_{\phi}=
-i {-\pc '(z-f(t,y)) \over \Delta(\ha)} {\de \over \de f}
+
\left[-{{\pc ' (z-f(t,y))\ha_k\langle i|\part_z|k\rangle}
\over \Delta(\ha)} +
\psi_i(z-f(t,y)) \right] {\de \over \de \ha_i} \; .
$$
Ordering ambiguities will change the transformation in subleading order
in $\hbar$.  These will not introduce anomalies because it is
known that it is possible to regulate the underlying $\phi^4$ theory and keep
Lorentz invariance.
So in terms of the field theory for $f$ and the components of $\xi$,
the Lorentz transformations
become
$$
\eqalign{
[M_{0y},f]&= i (t \part_y f + y \part_0 f) \cr
[M_{0y},\xi]&= i (t \part_y \xi + y \part_0 \xi) \cr
[M_{\mu z},f]&= i \bigg[-y_\mu +f \part_\mu f +
{\part_\mu f \over \Delta(\ha)} a_j\langle 0|z \part_z|j\rangle -
{\part_\mu a_j \over \Delta(\ha)} \langle0|z|j\rangle\bigg] \cr
[M_{\mu z},a_j]&= i\bigg[\part_\mu a_k\langle j|z|k\rangle
- \part_\mu a_i a_k
{\langle 0|z |i\rangle\langle j|\part_z|k\rangle\over \Delta(\ha)} \cr
&-\part_\mu f \left( \langle j|z|0\rangle +
\langle j|z \part_z|k\rangle a_k
-a_i a_k
{{\langle 0|z \part_z|k\rangle\langle j|\part_z|i\rangle}\over
\Delta(\ha)}\right)
+ f \part_\mu a_j\bigg] \; . \cr
}
$$
Again it is possible to get rid of $\xi$ by substituting its derivative
expansion
(good for low energies) in terms of $f$.  Since $\xi$ scales as
$\part^2$, the first terms depending on the massive modes $\xi$ in
the Lorentz transformations come in at order $\part^3$.  In the
action dependence upon $\xi$ enters first at order $\part^6$.
Thus the universal (independent of the potential) parts of the Lagrangian
for a theory with canonical kinetic term are
$$
{1 \over 2 \pi \alpha '} \left[1 - {1 \over 2} (\part f)^2 -
{1 \over 8} (\part f)^4 \right]
$$
and the corresponding universal parts of the Lorentz transformation which
are symmetries of this up to $O(\part^6)$ are
$$
[M_{\mu z}, f] = i [ - y_{\mu} + f \part_\mu f] \qquad
[M_{0y},f]= i (t \part_y f + y \part_0 f) \; .
$$
One can see that to this order the Lorentz generators form a
representation of the Lorentz algebra.  The irrelevant term in the
action proportional to $(\part f)^4$ gives an indication that $S$ is
proportional to area (the preceding lower order term is the action for
a free scalar field).  Since the Lorentz transformations
begin with a universal dimension $-1$ term, $y_\mu$ (for a standard
kinetic term for the underlying field theory),
there are cancellations between
renormalizable and nonrenormalizable terms in $S_{\rm eff}$ under the
symmetry group.  The nonrenormalizable terms reflect dependence
upon the short distance behavior of the underlying field theory.
These symmetries are up to order $\part^6$, $\hbar^2$ and boundary
terms.  One can also check explicitly the invariance of the measure
to this order.

The nonlinear transformations for the field $f$ are appropriate for
a Nambu-Goldstone boson.
One can turn the logic around and say that since the string breaks
translation invariance, the action for its Nambu-Goldstone
bosons is determined by the nonlinear transformations
under the broken symmetry.
Using the Volkov-Akulov
formalism,
[\hp] did this for the super Nielsen-Olesen vortex.

It is possible to compare this to the fundamental string in the light
cone gauge, where one imposes the conformal invariance constraints
$T_{++}= T_{--}= (\part_t X \pm \part_y X)^2 = 0$ on the
Polyakov string in conformal gauge.
One chooses $X^+ = x^+ + p^+ \tau$ and then uses
the constraint on $T_{\pm \pm}$
to solve for $X^-$ in terms of $X^2$.  With this choice the
complete action is
$ S = \int \beta (\part_\mu X^2)^2, $ which is renormalizable.
However, the conformal anomaly for dimension $D \ne 26$ means that
the constraints cannot be imposed, {\it i.e.},
solving for some of the degrees of freedom using the
constraints is inconsistent.  One way this shows up is
that the Lorentz transformations do not close in dimensions larger than 3.
In 3 dimensions, there is only one nonlinear Lorentz generator and there
is no problem with the closure of the algebra---one has to go to the
interacting string to see problems.

To compare directly to the induced string theory in
light cone gauge,
one would need to quantize the $\phi^4$ theory in light cone
gauge.  Quantization in light cone gauge is rather difficult to
do for interacting (non-integrable) scalar field theories.  The gauge
implied by the standard quantization of the
underlying $\phi^4$ field theory
is not light cone, as one can check explicitly.
Another comparison of light cone gauge fundamental  and effective strings
was made by Olesen[\ol].  He put unusual boundary conditions on Nambu-Goto
strings of length $R$ and showed that the Lorentz anomaly was
proportional to ${\alpha ' / R^2}$, which disappears as $R$ gets large.

The effective strings approximate structureless strings for
length scales $\gg m^{-1}$.  Besides the long wavelength
limit, one can make another comparison with structureless
strings, as Nielsen and Olesen did in their original paper
on flux tubes in the Abelian Higgs model[\no].
The solution $\pc(m, \sqrt{\lambda},z)$ depends on the parameters
$m, \lambda$ of the potential.  They asked the question: is it possible
to tune $\lambda$ so that fluctuations on the scale $m^{-1}$
don't excite the string width?  (They were looking for structureless strings
in $D \ne 26$.)
The energy of fluctuations associated with the field $f$ to leading order is
$ S \sim {1 \over \alpha '} \int (\part f)^2 $
where the tension $\alpha ' \sim [L^2]$, scales as length squared.
So for large $\alpha '$ fluctuations have low energy, and for small
$\alpha '$ fluctuations have large energy.  Asking the energy of fluctuations
to be too small to excite the internal structure of the string means
$ \sqrt{\alpha '} \gg m^{-1}  \; .  $
In terms of the model discussed in this talk, $1 / \alpha ' =
\int dz (\pc ')^2 \sim m^3/ \lambda$.  So the constraint becomes
$ \lambda \gg m \; .  $
Since the effective field theory was a derivative expansion assuming
$m$ large, taking $\lambda $ larger is very strong coupling--
analysis done in perturbation theory cannot be trusted anymore, and all
bets are off.  (For the Nielsen-Olesen vortex their constraint was
also strong coupling, $e \gg 1$.)  This is like the $\hbar \rightarrow
\infty$ limit, as they put it, so expanding about a classical solution
in a functional integral is meaningless.
For example the $\phi^4$ theory, when Euclidean,
is in the universality class of the high
temperature Ising model in this limit.

Before closing, it is worthwhile to mention some implicit strings
for contrast.  For these,
stringlike behavior is seen or expected but the strings are not fully
characterized.  One example is compact QED in $2+1$
dimensions.  Polyakov[\poly] demonstrated that there
is a linear potential between charges, which could be attributed to
a string.  Another
classic example is QCD, where string theory was first used.
At long distance, quarks and gluons are confined and one wants
to use variables providing a natural
description of the physics. Strings are a popular paradigm, but the
specific definition of a QCD string is unclear.  For perspective, we mention
three clues that suggest strings as a good description.

First of all, in the data Regge behavior was observed, {\it i.e},
for mesons and baryons, one found the relation $M^2 \sim J$
which one can model by a relativistic string (all the energy
comes from stretching it, $\rho = T$, density = tension), and
the ends move at the speed of light.  Then $J \propto MvR = ML$ and
$M= TL$, and so one gets the above mass/angular momentum relation.
This rule has been seen to be a good heuristic up to $J= 19/2$,
with the relation getting better and better at higher spin[\regge].

Secondly, strings were expected from lattice descriptions of QCD.
At strong coupling[\wils], one can rewrite the sum of gauge configurations as
the sum over surfaces.  Initially, it was thought that this rewriting
was in terms of noninteracting surfaces, but Weingarten showed
otherwise[\wein].  One could try to fix up
the noninteracting surface theory to account for the interactions in
various ways, for example by adding degrees of freedom, but the resulting
models became quite involved and hard to work with.

A third connection between strings and QCD was found in the
field theory description of QCD by 't Hooft[\largen].  In the large $N$ limit
of SU($N$) gauge theory (fixing ${\smash{g^2 N}}$), planar diagrams dominated
and could be viewed as tracing out the world sheet of a string (in group
space).  Large $N$ accounts for some aspects of phenomenology, so it might
be expected that the stringlike picture coming from large $N$ has
some validity.  In addition to the vast amount of work
done on QCD and strings several years ago, there have been a few
attempts to look at strings and QCD again.  I would like to
mention in particular the paper by Polchinski[\polqcd] and the review
[\polstring].  In [\dl] there is
a summary of a lot of the older data with the aim of isolating what
particularly stringy properties it implies.

In conclusion, this was a review of the quantization of effective strings,
in the case where there were explicit stringlike
solutions to the equation of motion.
The procedure was to rewrite the field theory in the one string
sector, introduce collective coordinates for the position of the string and
use equations of motion to substitute for the massive fields.  Only
physical degrees of freedom were present in the action.  The result was
a nonrenormalizable effective field theory on the string worldsheet,
which had terms to arbitrarily
high order in the derivative expansion, and leading order in $\hbar$.
The fundamental string's constraints had no relevance for the
effective string, which was not conformally invariant and
at short distances was no longer
a string.  For example, Lorentz transformations mixed
relevant and irrelevant terms in the action for the effective string.
It was possible to take the thin string limit by going to long wavelengths.
Naively, strong coupling would also give a thin string limit, but on
closer inspection, in this limit, all of the analysis has dubious validity.
\bigskip

J.D.C. thanks the organizers for a chance to lecture
and the students for their interesting comments and questions.
V.P. was supported by D.O.E. grant DE-FG02-90ER40542.
J.D.C. also thanks L. Brekke for pointing out reference [\ol].

\medskip

\refout

\end